\documentstyle[prl,aps,preprint,psfig,epsfig,floats]{revtex}
\begin{document}
\def\be{\begin{equation}}
\def\ee{\end{equation}}
\def\bc{\begin{center}}
\def\ec{\end{center}}
\def\bea{\begin{eqnarray}}
\def\eea{\end{eqnarray}}
\draft 
\title{Bose-Einstein condensation in  complex networks}
\author{Ginestra Bianconi$^{*}$ and Albert-L\'aszl\'o Barab\'asi$^{*,\ \dag}$ }
\address{$^{*}$Department of Physics, University of Notre Dame, Notre Dame,
IN 46556, USA \\ $^{\dag}$ Institute for Advanced Studies,
Collegium Budapest,
 Szenth\'aroms\'ag utca 2,\\ H-1014 Budapest,
Hungary }
\maketitle
\begin{abstract}
The evolution of many complex systems, including the world wide
web, business and citation networks is encoded in the dynamic web
describing the interactions between the system's constituents.
Despite their irreversible and non-equilibrium nature these
networks follow Bose statistics and can undergo Bose-Einstein
condensation. Addressing the dynamical properties of these
non-equilibrium systems within the framework of equilibrium
quantum gases predicts that the 'first-mover-advantage',
'fit-get-rich' and 'winner-takes-all' phenomena observed in
competitive systems are  thermodynamically distinct phases of the
underlying evolving networks.
\end{abstract}

\newpage
\narrowtext
Competition for links is a common feature of complex systems: on the 
world wide web sites compete for URLs to enhance their visibility $\cite{GL}$,
 in business 
world companies compete for links to consumers $\cite{Kirman}$ and in 
the scientific 
community scientists and publications compete for citations,
a measure of their impact on the field $\cite{Redner}$. A common feature of 
these systems is
 that the nodes self-organize into a complex network, whose topology and
 evolution closely reflects the dynamics and outcome of this competition
$\cite{GL,Redner,Sci,Comment,Nature}$. 
The reversibility of the microscopic processes that govern the evolution of
these 
systems, such as growth through the addition of new nodes, or the 
impossibility (citation network) and the high cost (business networks) 
to modify an established links give a distinctly far-from equilibrium character to the evolution of theses complex systems. Here we show that despite their
 non-equilibrium and irreversible nature, evolving networks can be 
mapped into an equilibrium Bose gas $\cite{Huang}$, nodes corresponding
 to energy 
levels and links representing particles. This mapping predicts that 
the common epithets used to characterize competitive systems, such 
as 'winner-takes-all', 'fit-get-rich', or 'first-mover-advantage' 
emerge naturally as thermodynamically and topologically distinct 
phases of the underlying complex evolving network. In particular, 
we predict that such networks can undergo Bose-Einstein condensation,
 in which a single node captures a macroscopic fraction of links.

{\it Fitness model--} Consider a network that grows through the
addition of new nodes such as the creation of new webpages,  the
emergence of new companies or the publication of new  papers.  At
each time step we add a new node, connecting it with $m$ links to
the nodes already present in the system. The  rate at which
 nodes acquire links can vary widely as supported by measurements on 
the www $\cite{Comment}$, and by empirical evidence in citation $\cite{Redner}$ and economic networks.
 To incorporate  the
different ability of the nodes to compete for links we
 assign a fitness parameter to each
node, $\eta$, chosen from a distribution $\rho(\eta)$, accounting
 for the differences in content of webpages, the quality of
products and marketing of companies, or the importance of the
findings reported in a publication. The
probability $\Pi_i$ that a new node  connects one of its $m$
links   to a node $i$ already present in the network depends on
the number of links, $k_i$, and
 on the fitness $\eta_i$ of  node $i$, such that
\be
\Pi_i=\frac{\eta_i k_i}{\sum_{\ell} \eta_{\ell} k_{\ell}}.
\label{p} \ee Equation $(\ref{p})$ incorporates in the simplest
possible way the fact that new nodes link preferentially to nodes
with higher $k$ $\cite{BAScience}$ (i.e. connecting to more
visible websites; favoring more established companies; or citing
more cited papers), and with larger fitness (i.e. websites with
better content;  companies with better products and sales practice; 
or papers with novel results). Thus fitness ($\eta_i$) and the
number of links ($k_i$) jointly determine the attractiveness and
evolution of a node.

\begin{figure}
\centerline{\epsfig{figure=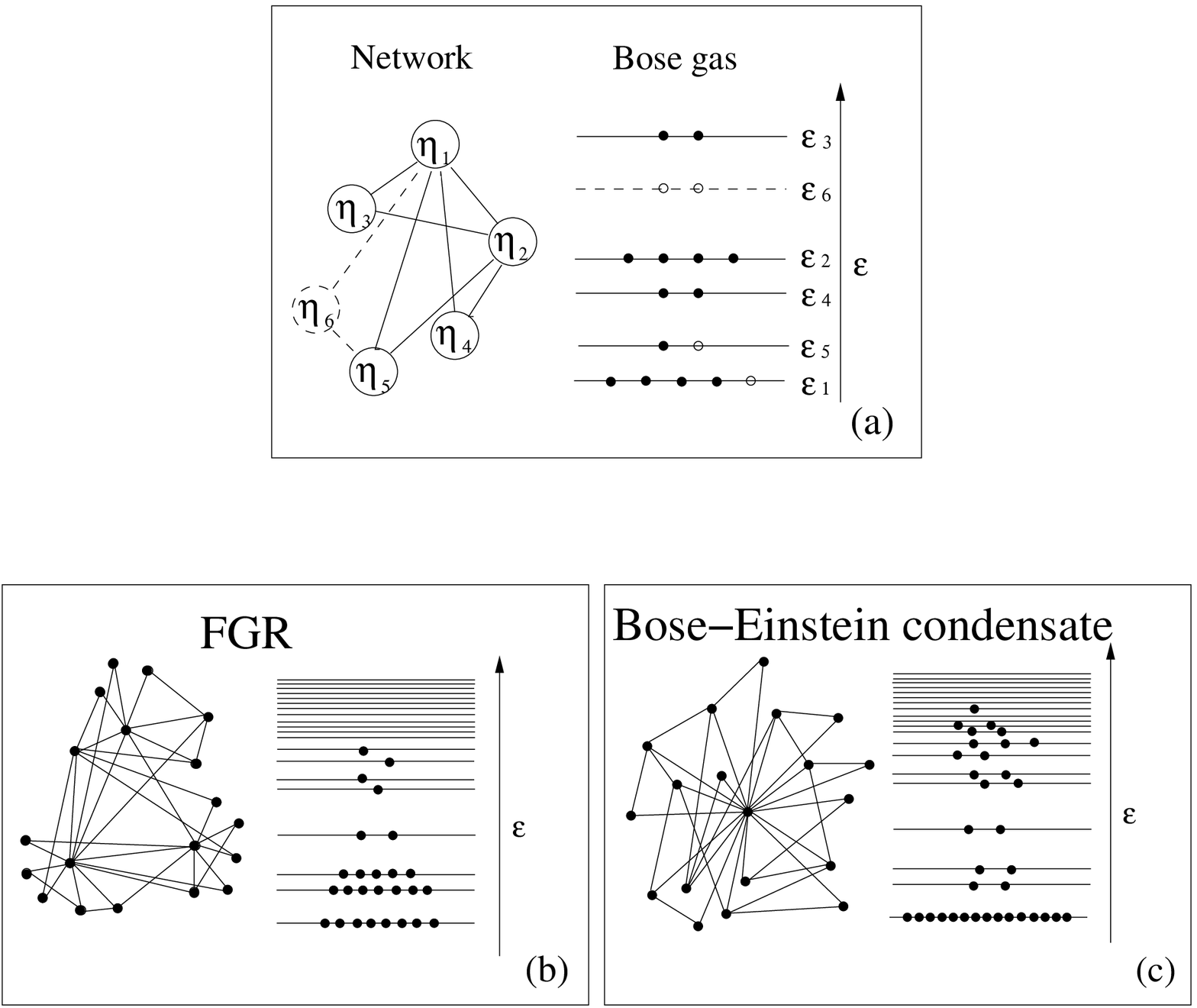,height=5.0in,width=6.0in,angle=0}}
\renewcommand\baselinestretch{1.00}
\caption{Schematic illustration of the mapping between the network model and
the Bose gas. ${\bf (a)}$ On the left we have a network of five nodes
 (continuous circles and lines), each
 characterized by a fitness, $\eta_i$, chosen randomly
from a distribution $\rho(\eta)$. Equation $(\ref{etaep})$ assigns
an energy $\epsilon_i$ to each $\eta_i$, generating a system  of
random energy levels (right). A link from node $i$ to node $j$
corresponds to a particle at  level $\epsilon_i$ and one at $\epsilon_j$.
 The network evolves by adding a new node (dashed
circle, $\eta_6$) at each timestep which connects to $m=2$ other
nodes (dashed lines), chosen randomly following $(\ref{p})$. In
the gas  this results in the addition of a new energy
level ($\epsilon_6$, dashed) populated by $m=2$ particles, and  the
deposition of $m=2$ other particles at energy levels to which
$\eta_6$ is connected to ($\epsilon_2$ and $\epsilon_5$). 
The number of energy levels and 
particles increase linearly with time, as $t$ and $2mt$,
respectively. {\bf (b)} In the FGR phase we have 
a continuous connectivity distribution,
the  several highly connected nodes 
linking the numerous small nodes together. In the energy diagram 
this corresponds to a decreasing occupation level with increasing energy.
{\bf (c)} In the Bose-Einstein condensate  
the fittest node attracts a finite fraction of all links, corresponding to 
a highly populated energy level, and only sparsely populated higher 
energies. In {\bf(b)} and {\bf(c)} the
 energy diagram shows only incoming links, 
ignoring the default $m=2$ particles on each energy level corresponding to
 the outgoing links to 
simplify the picture.} \label{fig1}
\end{figure}

 {\it Mapping to a Bose gas--}
Using a simple parameterization, we assign  an {\it energy} $\epsilon_i$ to
each node, determined by its fitness $\eta_i$ through the relation
\be \epsilon_i=-\frac{1}{\beta}\log \eta_i, \label{etaep}\ee where
$\beta$ is a parameter playing the role of inverse temperature,
 $\beta=1/T$, whose relevance to real networks will be discussed later.
  A link between two nodes   $i$ and $j$ 
with energy $\epsilon_i$ and $\epsilon_j$ (e.g. fitnesses 
$\eta_i$ and $\eta_j$) corresponds to  two non-interacting {\it
particles}  on the energy levels $\epsilon_i$ and $\epsilon_j$
(Fig.~$\ref{fig1}$). Adding a new node to the network corresponds to
 adding a new energy level
$\epsilon_i$ and $2m$ particles to the system. Of these $2m$ particles
  $m$
 are deposited on  the  level $\epsilon_i$  (corresponding
to the $m$ outgoing link  node $i$ has), while the other $m$
particles are distributed between the other energy levels
(representing the links pointing to $m$ nodes  present in the
system), the probability that a particle  lands on  level $i$
being given by $(\ref{p})$. Deposited particles are inert, i.e.
they are not allowed to jump to other energy levels.

Each node (energy level) added to the system at time $t_i$ with
energy $\epsilon_i$ is characterized by the occupation number
$k_i(t,\epsilon_i,t_i)$, denoting  the number of links (particles)
the node (energy level) has at time $t$.
 The rate at which   level $\epsilon_i$ acquires new
  particles is
 \be \frac{\partial k_i(\epsilon_{i},t,t_i)}{\partial t}= m \frac{e^{-\beta
\epsilon}\ k_i(\epsilon_{i},t,t_i)} {Z_t}, \label{continuum2} \ee
where  $Z_t$ is  the partition function,  defined as \be
Z_t=\sum_{j=1}^{t} e^{-\beta \epsilon_j} k_j(\epsilon_j,t,t_j).
\label{zts}\ee We assume that each node increases its connectivity
following a power-law (we demonstrate the self-consistency of this
assumption later)
\be
k_i(\epsilon_{i},t,t_i)= m \left( \frac{t}{t_i}
\right)^{f(\epsilon_{i})}, \label{power}\ee where $f(\epsilon)$ is
the energy dependent dynamic exponent.
 Since
$\eta$ is chosen randomly from the distribution $\rho(\eta)$, the
energy levels are chosen  from the distribution $g(\epsilon)=\beta
\rho(e^{-\beta \epsilon})e^{-\beta \epsilon}$. We can now
determine  $Z_t$ by averaging over $g(\epsilon)$, i.e.
 \bea <Z_t >&=&
\int d\epsilon g(\epsilon)\ \int_1^t dt_{0} e^{-\beta \epsilon}
k(\epsilon,t,t_0) \nonumber \\  &=&m \ z^{-1} t\
(1+O(t^{-\alpha})), \label{lim}\eea where \bea
 \frac{1}{z}=\int {d\epsilon
g(\epsilon) \frac{ e^{- \beta \epsilon}}{1-f(\epsilon)}}
\label{C1} \eea   is the inverse fugacity. Since $z$ is positive
for any $\beta\ne 0$ we introduce the chemical potential, $\mu$,
as $z=e^{\beta \mu}$, which allow us to write ($\ref{lim}$) and
($\ref{C1}$) as
\be
 e^{-\beta\mu}=\lim_{t \rightarrow \infty} \frac{<Z_t>}{t}.\label{mu}\ee
Using ($\ref{mu}$)  we can solve the continuum equation
($\ref{continuum2}$) finding in a self-consistent way solutions of
form $(\ref{power})$, where the dynamic exponent is
\be
f(\epsilon)=e^{-\beta (\epsilon-\mu)}. \label{f} \ee  Combining
($\ref{C1}$) and ($\ref{f}$), we find that the chemical potential
is the solution of the equation
\be
I(\beta,\mu)=\int d\epsilon g(\epsilon) \frac{ 1}{e^{\beta
(\epsilon-\mu)}-1}=1. \label{selfc} \ee The system defined above
has a number of properties that make it an unlikely candidate for
an equilibrium Bose gas $\cite{Huang}$. First, the inertness of
the particles is a non-equilibrium feature, in contrast with the
ability of particles in a quantum gas to jump between energy levels, 
leading to a temperature driven equilibration.
Second, both the number of eligible energy levels (nodes) and
 particles populating them (links) increase linearly
in time, in contrast with the fixed system size employed in quantum 
systems.
  Despite these apparent conflicts, Eq.~$(\ref{selfc})$ indicates that in the
thermodynamic limit ($t \rightarrow \infty$) the fitness model maps
 into a Bose
gas. Indeed, since in an ideal gas of volume $v=1$ we have
$\cite{Huang}$
\be
\int d\epsilon g(\epsilon) n(\epsilon)=1, \ee where $n(\epsilon)$
is the occupation number of a level with energy $\epsilon$
Equation~$(\ref{selfc})$ indicates that for the inert gas inspired by
the fitness model the  occupation number follows the familiar Bose
statistics $\cite{Huang}$
 \be n(\epsilon)=\frac{ 1}{e^{\beta
(\epsilon-\mu)}-1}, \label{nbose}\ee
i.e. the evolving network maps into a Bose gas.
Thus the irreversebility and the inertness of the network are resolved by the stationarity of the asymptotic 
distribution, allowing the occupation number 
to follow Bose statistics in the thermodynamic 
limit $t\rightarrow \infty$.

{\it Bose-Einstein condensation--} The solutions
$(\ref{power})$, $(\ref{lim})$ and $(\ref{f})$  exist only
when there is a $\mu$ that satisfies Eq.~$(\ref{selfc})$. However,
   $I(\beta,\mu)$
 defined in $(\ref{selfc})$  takes its
maximum at $\mu=0$, thus when $I(\beta,0)<1$ for a given $\beta$
and $g(\epsilon)$, Eq.~$(\ref{selfc})$ has no solution. The
absence of a solution is a well-known signature of Bose-Einstein
condensation $\cite{Huang}$, indicating that  a finite
$n_0(\beta)$ fraction of particles condensate on the  lowest
energy level. Indeed, due to mass conservation at  time $t$ we
have
  $t$ energy levels populated by $2mt$ particles, i.e.
\be
2mt=\sum_{t_0=1}^{t} k(\epsilon_{t_0},t,t_0)=mt+mtI(\beta,\mu).
\label{massc}\ee  When $I(\beta,0)<1$, Eq.~$(\ref{massc})$  has to
be replaced with
\be
2mt=mt+mtI(\beta,\mu)+n_0(\beta), \ee where $n_0(\beta)$ is given
by $\cite{Huang}$
\be
\frac{n_0(\beta)}{mt}=1-I(\beta,0). \label{frac}\ee The occupancy
of the lowest energy level   corresponds to the number of links
the node with the largest fitness has. Thus the emergence of a
nonzero $n_0(\beta)$, a signature of  Bose-Einstein condensation
in quantum gases,  represents   a 'winner-takes-all' phenomena for
 networks, the fittest node acquiring a finite fraction of the
links, independent of the size of the network.

 The mapping to a Bose gas and the possibility of  Bose-Einstein
condensation in random networks predicts the existence of three
distinct phases  characterizing the dynamical properties of
evolving networks: (a) a scale-free phase, (b) a fit-get-rich
phase and (c) a Bose-Einstein condensate. Next we discuss 
each of these possible phase separately.

 {\it (a) Scale-free phase--} When all nodes have
the same fitness, i.e. $\rho(\eta)=\delta(\eta-1)$,
($g(\epsilon)=\delta(\epsilon)$), the  model reduces to the
scale-free model $\cite{BAScience}$, introduced to account for the
power-law connectivity distribution observed in diverse systems,
 such as the www $\cite{Sci,Nature}$, actor network
$\cite{WS,BAScience}$, Internet $\cite{Faloutsos}$ or citation
networks $\cite{Redner}$. The model describes a 'first-mover-wins'
behavior, in which the
 oldest nodes  acquire  most links. Indeed,
$(\ref{f})$ predicts that $f(\epsilon)=1/2$, i.e. according to
$(\ref{power})$ all nodes increase their connectivity as
$t^{1/2}$, the older nodes with  smaller $t_i$ having larger
$k_i$. However, the oldest and 'richest' node is not an absolute
winner, since its share of  links, $k_{max}(t)/(mt)$, decays to zero as
$t^{-1/2}$ in the thermodynamic limit. Thus  a continuous
hierarchy of large  nodes coexist, such that the
connectivity distribution $P(k)$, giving the probability to have a
node with $k$ links, follows a power law $P(k)\sim k^{-3}$
$\cite{BAScience,Mend}$. Rewiring, aging, and other local processes  can modify the scaling exponents or introduce exponential 
cutoffs in $P(k)$ $\cite{Mend,Red_new,Amacd}$ while leaving 
the thermodynamic character of the phase unchanged.

 {\it (b) Fit-get-rich (FGR) phase--} This phase emerges in
systems for which   nodes  have different fitnesses and
Eq.~$(\ref{selfc})$ has a solution (i.e. $I(\beta,\mu)=1$).
Equation $(\ref{power})$ indicates that each node increases its
connectivity in time, but the dynamic exponent depends on the fitness, being
 larger for nodes with
higher fitness. This  allows fitter
nodes to join the system at a later time and  surpass the less
 fit but older nodes by acquiring links at
higher rate $\cite{Comment}$. Consequently,
this phase predicts a 'fit-get-rich' phenomena, in which, with time, the
fitter   prevails. But, while there is a clear winner, similar
to the scale-free phase the fittest node's  share of all links
decreases to zero in the thermodynamic limit. Indeed, since
$f(\epsilon)<1$, the relative connectivity of the fittest node
 decrease  as $k(\epsilon_{min},t)/(mt)\sim t^{f(\epsilon_{min})-1}$.
 This competition again leads to the emergence of a
hierarchy of a few large 'hubs' accompanied by  many less
connected nodes,
 $P(k)$  following a power-law
$P(k)\sim k^{-\gamma}$, where $\gamma$ can be calculated 
analytically if  $\rho(\eta)$ is known.

 {\it (c) Bose-Einstein (BE) condensate --}
   Bose-Einstein condensation
appears when $(\ref{selfc})$ has no solution, at which point
$(\ref{power})$, $(\ref{f})$, $(\ref{selfc})$ break down. In the
competition for links the node with the largest fitness emerges as
a clear winner, a finite fraction of particles $(n_0(\beta))$
landing on this energy level. Thus BE condensation predicts a real
'winner-takes-all' phenomena, in which the fittest node is not only the 
largest, but despite the
continuous emergence of new nodes that compete for links, it
always acquires a finite fraction of  links (Eq.~$(\ref{frac})$).

 To demonstrate the existence of a phase
 transition from the FGR phase to a
 BE condensate in a network, we  consider
 a formally simple case, assuming that the energy (fitness)
 distribution follows
\be
g(\epsilon)=C{\epsilon}^{\theta},
\label{pe}
\ee where $\theta$ is a free parameter and the
energies are chosen from $\epsilon \in (0,\epsilon_{max})$, normalization 
giving $C=(\theta+1)/{\epsilon^{\theta+1}_{max}}$.
For this class of distributions the condition for a Bose condensation is
\be
\frac{\theta+1}{{(\beta \epsilon_{max})}^{\theta+1}} \int^{\beta \epsilon_{max}}_{\beta \epsilon_{min}(t)}dx \frac{x^\theta}{e^x-1}<1,
\ee
where $\epsilon_{min}(t)$ corresponds to the lowest energy (fittest) node present in the system at time $t$.
Extending the limits of integration to zero and to infinity, we find 
the lower bound for the critical temperature $T_{BE}=1/\beta_{BE}$
\be
T_{BE}>\epsilon_{max}{\left(\zeta(\theta+1)\Gamma(\theta+2)\right)}^{-1/(\theta+1)}.
\label{tcteorico}
\ee
To demonstrate the emergence of Bose-Einstein condensation
 near the predicted $T_{BE}$, we 
simulated numerically the  discrete network model described above, using 
the energy distribution  $(\ref{pe})$. The chemical potential, $\mu$, 
measured numerically indicates a sharp transition from a positive 
to a negative value (Fig. 2a), 
corresponding to the predicted phase transition between the BE and the FGR
 phases. The difference between the network dynamics in the two phases is 
illustrated in Fig.2b, where we plot the relative occupation number of the 
most connected node for different temperatures. We find that the ratio 
$k_{max}(t)/m t$ is independent on time in the BE phase, 
indicating that  
the largest node maintains a finite fraction of the total number of 
links even as the network continues to expand,  a signature of BE
 condensation. In contrast, for $T>T_{BE}$, 
the most connected node gradually looses its  share of links,
 $k_{max}(t)/m t$ 
decreasing continuously with time. 
The numerically determined phase diagram (Fig. 2b)  confirms that  
the analytical prediction $(\ref{tcteorico})$ offers
 a lower bound for  $T_{BE}$.
 To predict the precise value of $T_{BE}$ one needs to 
 incorporate the interplay between
 the convergence of 
$\epsilon_{min}(t)$ to zero and the thermodynamics of BE condensation.

Since real networks have a fixed ($T$ independent) $\rho(\eta)$ fitness
distribution, whether they are in the BE or FGR phase is independent of
$T$. Indeed, for $\rho(\eta)~(1-\eta)^{\lambda}$ the network undergoes a BE
condensation for $\lambda>\lambda_{BE}=1$, thus $T$, introduced during the
calculations, plays the role of a dummy variable that at the end
vanishes from all topologically relevant quantities. The existence of
$T_{BE}$ in the numerically studied model (Fig. 2) is rooted in our
technically simpler choice of defining $g(\epsilon)$ to be independent of
$T$, providing  a richer phase space to demonstrate the existence of a
phase transition in a discrete network model. However, as the inset in
Fig. 2b shows, by changing $\theta$ the phase transition emerges for fixed
$T$ as well.

Do the www,  business or  citation networks represent a
Bose-Einstein condensate, or are described by the FGR phase
without a dominating winner?
It is well known that in certain markets some companies
did gain and maintain an unusually high market share. A
much publicized case is the continued dominance of  Microsoft
 in the rapidly expanding  operating systems market,
 indicating that  business
networks do develop 'winner-takes-all' phenomena, similar to
a Bose-Einstein condensate. The situation
of the www is more complex:  while indeed some webpages did
capture an unusually high number of links, it is less clear
weather  they maintain a finite fraction of all links as the www
grows (as expected if the www is a BE condensate), or will loose
market share, leading to the coexistence of a continuous hierarchy
of large websites,  a signature of the FGR phase. While the power-law 
connectivity distribution holding over six order of magnitude
support the FGR scale-free phase $\cite{Nature,IBM}$, the most
recent study involving more that 200 million nodes did identify a
run-away node of approximately $10^7$ links $\cite{IBM}$, an
apparently clear and dominating winner.  Since $\rho(\eta)$ could be explicitly
determined from network data, the phase to which various networks belong 
to could be decided in
the future. Uncovering the prevailing behavior in these and other
complex evolving networks (e.g. social networks $\cite{Social}$,
 ecological and transportation  webs $\cite{Martinez,Ba}$, etc.)
is a formidable task that will require careful quantitative
studies on real networks. The identification of the possible
phases offered in this paper might provide the appropriate
analytical framework for such studies.


\newpage

\begin{figure}
\centerline{\epsfig{figure=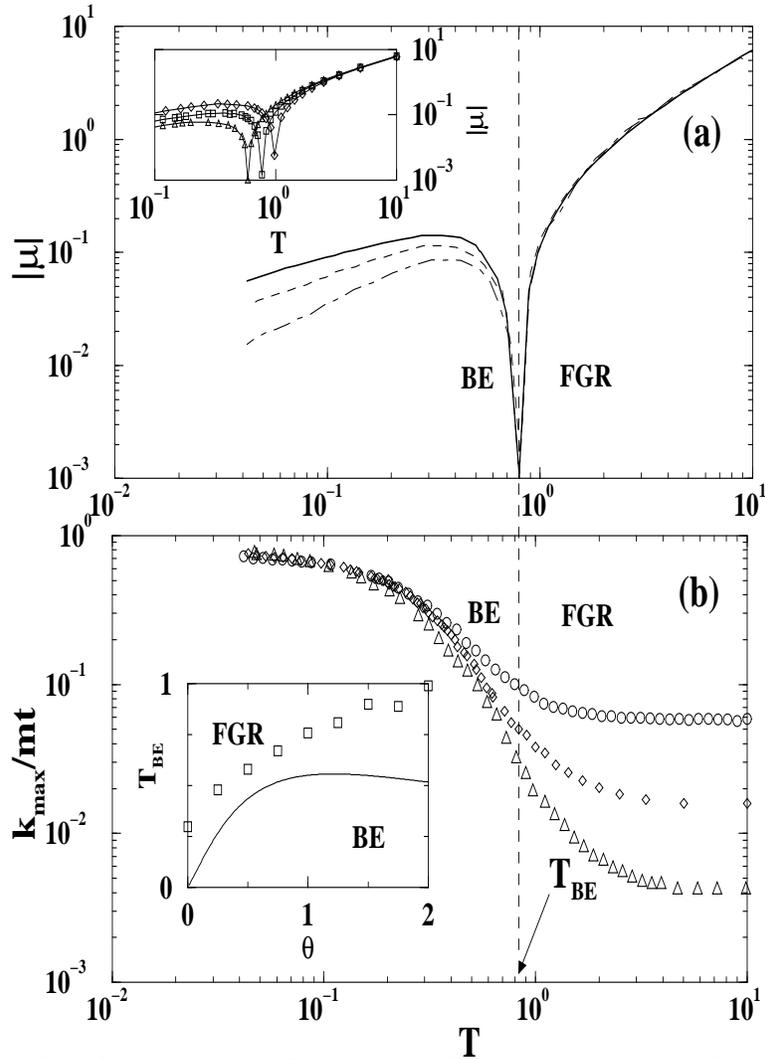,height=5.5in,width=4in,angle=0}}
\renewcommand\baselinestretch{1.00}
\caption{Numerical evidence for
Bose-Einstein condensation in a network model. {\bf(a)}
Choosing the energies from the distribution $(\ref{pe})$ with
 $\epsilon_{max}=1$, we calculated the chemical potential
$\mu$ numerically as the network evolved in time, using $(\ref{mu})$.
Due to the wide range over which $\mu$ varies, we plot $|\mu |$ on 
a logarithmic scale. 
  The temperature  at which $\mu$ changes sign corresponds to the sharp drop
in $|\mu|$ 
on the figure, and identifies  the
critical temperature
$T_{BE}$ for  Bose-Einstein condensation.
The data are shown for $\theta=1$ and for different times (i.e. system sizes)
$t=10^3$ (continuous), $10^4$ (dashed), $10^5$ (long-dashed),
 being averaged over $500, 100$ and $30$ runs respectively. The inset shows the chemical potential for different 
values of the exponent $\theta$ in $(\ref{pe})$, i.e. $\theta=0.5,1.0,2.0$,
indicating the $\theta$ dependence of $T_{BE}$.
{\bf (b)}
Fraction of the total number of links
connected to the most connected ('winner') node, $k_{max}/(mt)$, plotted as a function of $T$,
 shown for  $m=2$ and $\theta=1$. 
The three curves, corresponding to different system sizes recorded at 
 $t=10^3,10^4,10^5$ indicate the difference between the two phases: 
in the BE phase (left) the fittest node maintains a
 finite fraction of links even
 as the system expands, while in the FGR phase (right) the fraction of links 
connected to the most connected node decreases with time.
The inset shows the $(\theta,T_{BE})$ phase diagram, the continuous line
 corresponding to the lower bound  predicted by Eq.~$(\ref{tcteorico})$,
 while the symbols represent the numerically measured $T_{BE}$ as indicated by
the position of the peaks in Inset(a).
} 
\label{fig2}
\end{figure}
\end{document}